\documentclass[]{spie}  

\usepackage{amsmath,amsfonts,amssymb}
\usepackage{graphicx}
\usepackage[colorlinks=true, allcolors=blue]{hyperref}
\usepackage{enumerate}

\usepackage{upgreek}

\usepackage{acronym}
\acrodef{AO}[AO]{Adaptive Optics}
\acrodef{DM}[DM]{Deformable Mirror}
\acrodef{FWHM}[FWHM]{Full Width Half Maximum}
\acrodef{WFS}[WFS]{Wavefront Sensor}
\acrodef{SHWFS}[SHWFS]{Shack-Hartmann Wavefront Sensor}
\acrodef{PSF}[PSF]{Point Spread Function}
\acrodef{NIR}[NIR]{Near Infrared}
\acrodef{LGS}[LGS]{Laser Guide Star}
\acrodef{NGS}[NGS]{Natural Guide Star}
\acrodef{ELT}[ELT]{Extremely Large Telescope}
\acrodef{MM}[MM]{multimode}
\acrodef{SM}[SM]{single-mode}
\acrodef{PL}[PL]{photonic lantern}
\acrodef{SNR}[SNR]{signal to noise ratio}

\title{Demonstration of a photonic lantern low order wavefront sensor using an adaptive optics testbed}

\author[a]{Mark K. Corrigan}
\author[a]{Timothy J. Morris}
\author[b]{Robert J. Harris}
\author[b]{Theodoros Anagnos}
\affil[a]{Department of Physics, Durham University, South Road, Durham, DH1 3LE, UK}
\affil[b]{Landessternwarte, Zentrum f\"ur Astronomie der Universit\"at Heidelberg, K\"{o}nigstuhl 12, 69117 Heidelberg, Germany}

\authorinfo{Further author information: (Send correspondence to Mark K. Corrigan)\\E-mail: m.k.k.corrigan@durham.ac.uk}

\begin{document} 
\maketitle

\begin{abstract}

We demonstrate the use of an optimized 5 core photonic lantern (PL) to simultaneously measure tip/tilt errors at the telescope focal plane, while also providing the input to an instrument. By replacing a single mode (SM) fiber with the PL we show that it is possible to stabilize the input PSF to an instrument due to non-common path tip/tilt aberrations in an adaptive optics system. We show the PL in two different regimes, (i) using only the outer cores for tip/tilt measurements while feeding an instrument with the central core and, (ii) using all cores to measure tip/tilt when used in an instrument such as a spectrograph. In simulations our PL displays the ability to retrieve tip/tilt measurements in a linear range of $\pm 55$milliarcseconds. At the designed central wavelength of $1.55\upmu$m, configuration (i) matches the throughput of an on-axis SM fiber but declines as we move away from this wavelength. In configuration (ii) we make use of the whole multimode input of the PL resulting in a potential increase of overall throughput compared to a SM fiber, while eliminating modal noise.
\end{abstract}


\section{Introduction}
\label{sec:intro}  

The \ac{PL} is a photonic device becoming increasingly popular within the fields of astrophotonics and telecommunications. \ac{PL}s have previously been demonstrated to efficiently evolve modes from a \ac{MM} waveguide into many individual \ac{SM} waveguides \cite{Birks:15}. If light can be efficiently coupled to the SM waveguides of a PL then several aspects of astronomical instrument design and performance can be improved, including elimination of modal noise \cite{doi:10.1093/mnras/stv410}, a reduction in the sky background \cite{Birks:12}, and a reduction in the overall spectrograph size \cite{2012arXiv1212.4867S}.

PLs can also be particularly useful for high resolution spectroscopy. Typically,  MM fibers are used to achieve a high instrument throughput with the limitation that spectral resolving power is then deteriorated by fiber modal noise \cite{doi:10.1111/j.1365-2966.2011.19312.x}. Coupling to the SM waveguides can be optimized through the use of extreme \ac{AO} to couple a very high Strehl ratio Point Spread Function (PSF) directly into SM fibers \cite{doi:10.1117/12.316767,doi:10.1117/12.2234299, doi:10.1117/12.2233135}. To achieve maximum coupling into a SM fiber after an Adaptive Optics (AO) system, the elimination of non-common path (NCP) aberrations between the wavefront sensor and the science focal plane is essential. A large source of these NCP errors can be attributed to imperfect correction of telescope vibrations \cite{doi:10.1117/12.925984}, misalignment, or differential aberrations caused by the optical components between the wavefront sensor (WFS) and science focal plane. These can also vary slowly throughout observations due to a variety of reasons such as gravitational flexures or thermal variations.

\begin{figure}[!h]
\hspace*{-1cm}
\centering\includegraphics[width=18cm]{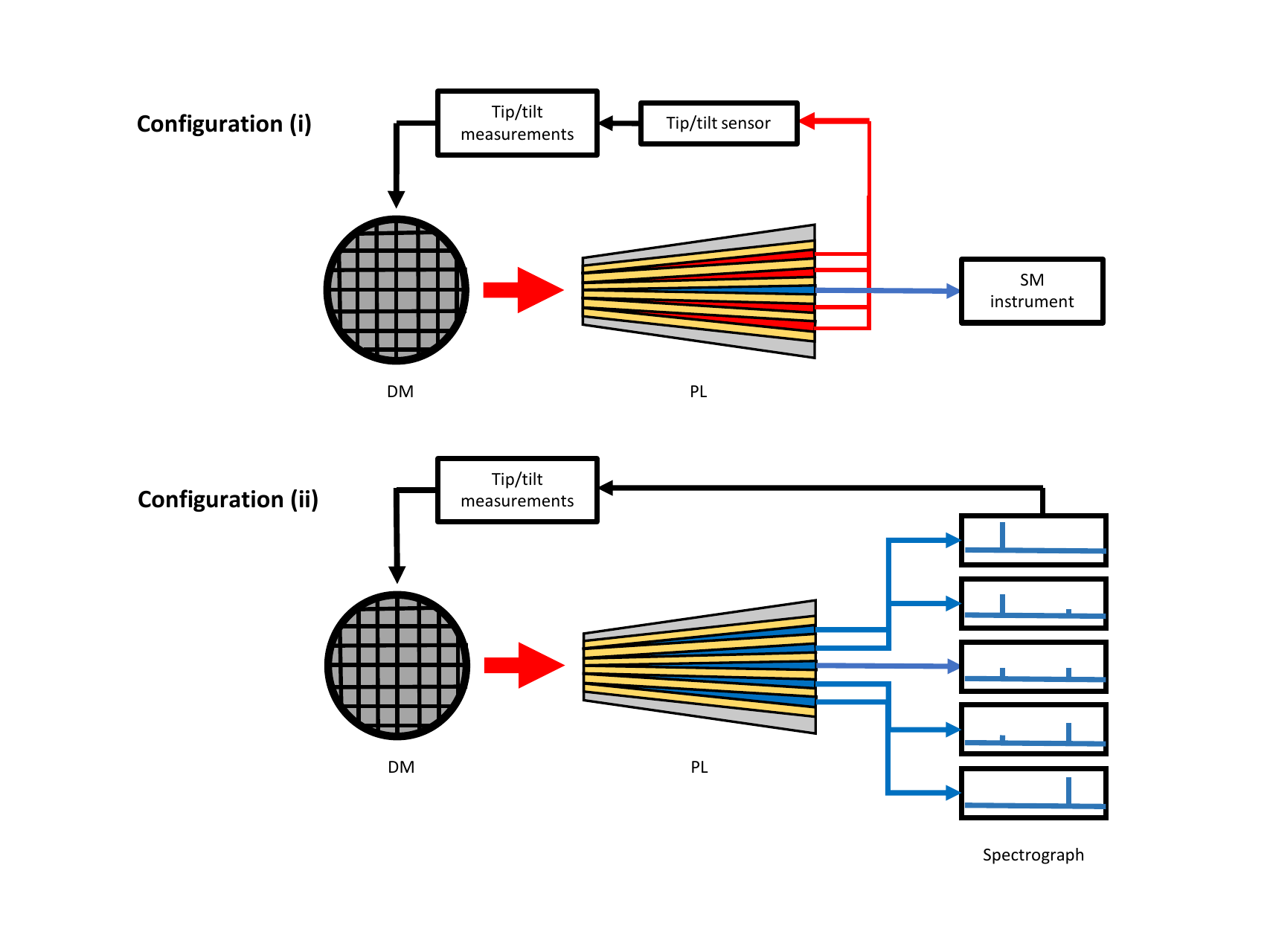}
\caption {Diagram of the two configurations in which the PL can be used in an AO system. Configuration (i) uses light from the outer cores to make tip/tilt measurements while supplying an instrument with light from the central core. (ii) uses spectra from all cores to make long exposure tip/tilt measurements.} \label{config_diagram}
\end{figure}

We have previously shown that in simulation a four core PL can be used as a focal plane tip/tilt sensor which behaves similarly to a quad-cell\cite{doi:10.1117/12.2230568}. Alternatively, K. Harrington et al.\cite{doi:10.1117/12.2306458} has presented a 3 core PL as a tip/tilt WFS placed in the pupil plane. Expanding on these studies, we present a PL design with simulation results which show replacing a SM fiber with a PL as a feed to an instrument can stabilize low-order aberrations, primarily tip/tilt, and thus increase the overall instrument fiber transmission. We introduce an AO fed fiber based PL in two configurations (shown in Figure~\ref{config_diagram}):

\begin{enumerate}[(i)]
\item Making use of a central SM output to supply light to an instrument, while also making residual tip/tilt measurements in the telescope focal plane using the outer SM cores of the PL. 

\item Encompassing the spectra from all cores to provide correction for slowly-varying errors. The relative intensities at particular wavelengths are analyzed and a measurement of the tip/tilt made.
\end{enumerate}

Both configurations have the benefit of utilizing all the incoming light needed for the instrument without sacrificing anything on the WFS. 

In this paper we introduce a throughput optimized fiber-based PL with 5 SM cores (4 outer cores and 1 central) in section \ref{lantern_design}. We show the simulation results of the PL's throughput compared to a single SM fiber in section \ref{wavelength} and the tip/tilt determination compared to a quad-cell in section \ref{tip/tilt}. Finally in section \ref{experiment} we present the optical design of a future experiment to test the PL in the lab. 


The data presented in this report was simulated using the RSoft CAD program with the BeamPROP add-on software \cite{rsoft}.

\section{Photonic lantern design} \label{lantern_design}

\begin{figure}[!h]

\centering\includegraphics[width=13cm]{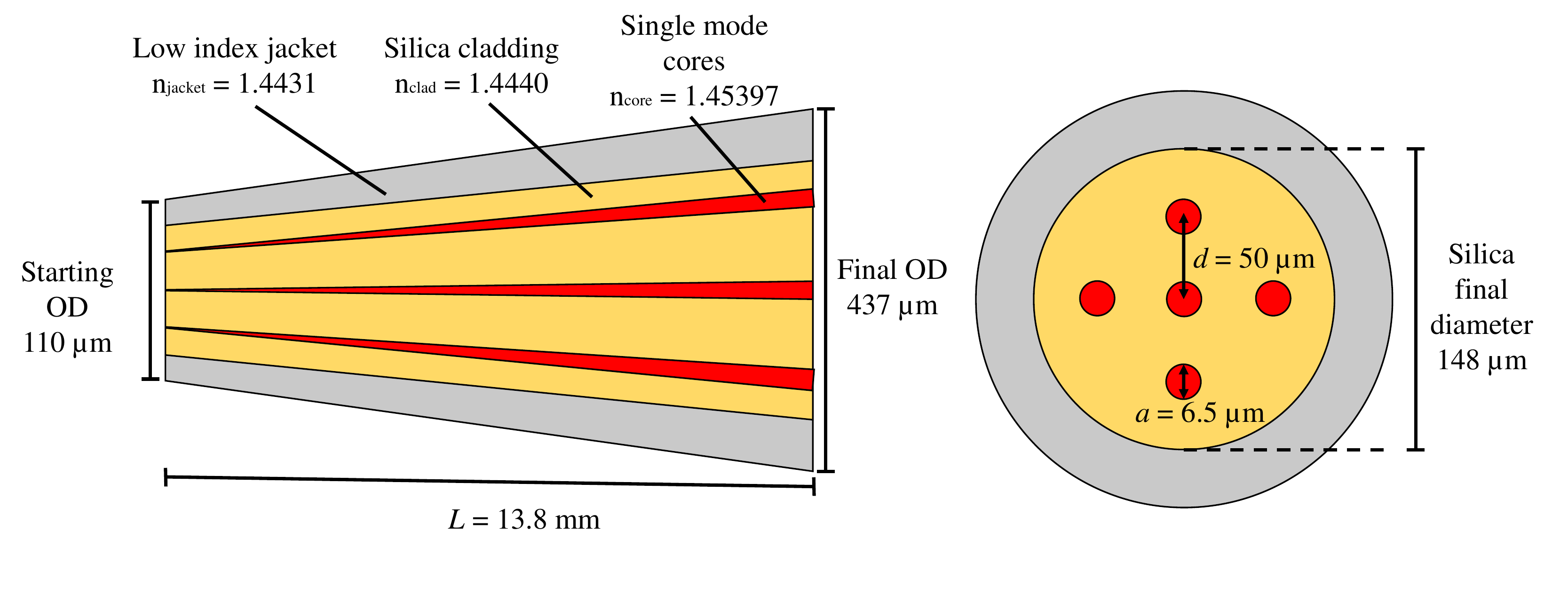}
\caption{Schematic of the PL used in simulations. Where L is the overall length of the taper region, a is the SM core diameter and d is the final separation between the cores.} \label{lantern_schematic}
\end{figure}

Our simulations consist of a fiber PL, with the initial specifications based on a previously designed and manufactured PL \cite{Noordegraaf:09,Leon-Saval:10}. We chose this PL as it has been previously studied and, using current techniques, can be realistically produced with the desired parameters. This PL consisted of 7 SM cores embedded in a silica cladding, with a low index jacket surrounding this structure. We have changed the structure to a 5 core PL to reduce the number of detector pixels required to measure the tip/tilt to 4, comparable to a quad-cell, as seen in Figure~\ref{lantern_schematic}. The parameters of the end face of the PL (right image of Figure~\ref{lantern_schematic}) are shown in Table~\ref{param_tab}. The end face structure was scaled down linearly to a starting outer diameter of $110\upmu$m at the MM end of the taper.

In our simulations we removed the MM fiber section which typically precedes the taper in most PLs, as beating between modes within this section leads to wavelength dependent outputs\cite{Cassidy:85} which was investigated also in \cite{doi:10.1117/12.2230568}. However, these effects are still present along the taper region before the cores separate out into the uncoupled 5 SM waveguides and is detailed further in sections \ref{optimisation} and \ref{wavelength}. 

\subsection{Optimization} \label{optimisation}

To optimize the PL design, we first maximized the throughput of the PL's central core by optimizing the Full Width Half Maximum (FWHM) of the PSF coupling into the lantern. By using the MOST tool within RSoft, we scanned the FWHM values of the input PSF for the PL and the SM fiber at a wavelength of $1.55\upmu$m. The maximum throughput from the PL's output central core was achieved for a PSF with a FWHM of $19.3\upmu$m. We then compared the PL to a SM fiber with parameters shown in Table \ref{param_tab} to match the PL's central SM core. For the individual SM fiber a PSF FWHM of $8.71\upmu$m was found to provide optimal throughput. This implies that for observations, at a given wavelength and telescope aperture, the input f-ratio at the lantern for optimal coupling would be a factor of 2.22 slower than that required for the SM fiber. The magnitude of image motion at the focal plane is preserved between SM and PL by expressing image motion in terms of the fiber radius. Therefore the equivalent focal plane image motion of one SM core radius ($3.25\upmu$m) on the SM fiber input would result in $7.20\upmu$m on the PL input.

Once the optimal input PSF f-ratio is known, we can start to optimize the interaction length of the PL (the area in which the PL cores are still coupled). This can be achieved by altering multiple parameters, such as core separation and taper geometry. Altering these parameters results in a lengthening or shortening of the interaction length which results in changes of the chromatic beating effects investigated previously\cite{doi:10.1117/12.2230568}. For the PL used in our simulations we optimized the overall taper length to retrieve the maximum throughput from the central core at $1.55\upmu$m. We found that for the fixed core separation we have chosen, the optimal taper length was 13.8mm.

\begin{table}
\centering
 \begin{tabular}{|c||c| c|} 
 \hline
 \textbf{Parameters} & \textbf{PL} & \textbf{SM Fiber} \\
 \hline
 n$\mathrm{_{core}}$ & 1.45397 & 1.45397 \\
 n$\mathrm{_{clad}} $ & 1.4440 & 1.4440 \\
 n$\mathrm{_{jacket}} $ & 1.4431 & - \\
 Length, $L$ (mm) & 13.8 & 13.8 \\
 SM core separation, $d$ ($\upmu$m) & 50 & - \\
 SM core diameter, $a$ ($\upmu$m) & 6.5 & 6.5 \\
 Cladding diameter ($\upmu$m) & 148 & - \\
 Jacket diameter ($\upmu$m) & 437 & - \\
 Input PSF FWHM ($\upmu$m) & 19.3 & 8.71 \\
 Input f-ratio (4.2m telescope @ 1.55$\upmu$m wavelength) & 9.50 & 4.29 \\
 \hline
\end{tabular}
\vspace*{0.5cm}
\caption{Simulation design parameters used for the PL and the SM fiber.}\label{param_tab}
\end{table}

\section{Simulation Design and Results}
\label{simulation}

\subsection{Wavelength dependence} \label{wavelength}

For the device to be useful for applications in a variety of instruments, we require the widest possible range of working wavelengths. The coupling conditions into an optical waveguide are governed by the PSF and the mode field diameters, which are wavelength dependent. Furthermore, output of the central core of the PL is affected by the modal coupling conditions along the PL taper, which are also wavelength dependent. Therefore, for configuration (i) the central core of the PL should have a limited range of working wavelengths compared to that of a SM waveguide, for a given PL transition (i.e. the designed core parameters and transition length). 

\begin{figure}[!h]
\centering\includegraphics[width=14cm]{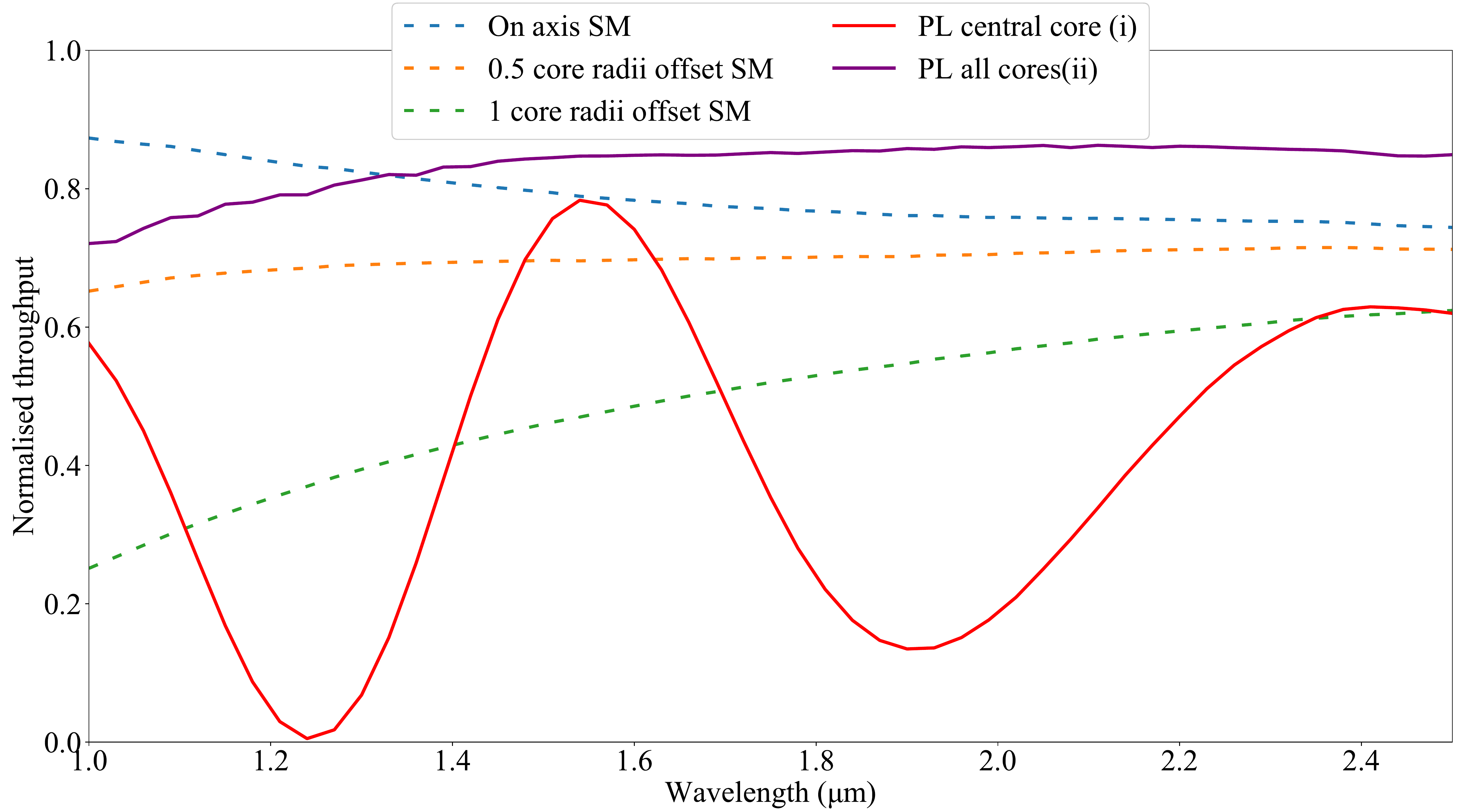}
\caption{Normalized throughput vs. wavelength for an ideal SM fiber and the PL, where we have ignored material effects. We have displayed the throughput of a SM fiber (dashed lines) for varying amounts of offset of the input. We compare against the case where we can correct for the image motion on the PL (solid lines) such that it is on-axis.} \label{throughput_plots}
\end{figure}

Figure~\ref{throughput_plots} shows the wavelength dependence on throughput (normalized to the total input field) for the SM fiber, the central core of the PL (i) and the total throughput of all cores (ii) of the PL. We have ignored material effects, such as attenuation, in our simulations to explore the wavelength behavior over a wide range. The on-axis case of the PL was compared to different offset values on the input to the SM fiber to simulate the image motion. This way we can estimate the level of correction required over a certain wavelength range to argue that using a PL is preferable to using a SM fiber. 

When using the PL in configuration (i), we can see that at a central wavelength of $1.55\upmu$m the on-axis central core of the PL can match the 78$\%$ throughput of a SM fiber and better throughput than a SM fiber when the input to the fiber is offset by a single SM core radius. However, the throughput is very dependent on wavelength and this PL will only outperform the SM fiber over the wavelength range of $1.48$$-$$1.62\upmu$m for a 0.5 core radii offset of the input and between $1.40$$-$$1.69\upmu$m for a 1 core radii offset. The throughput response of the SM fiber is also much flatter than the PL over a large wavelength range, which has been previously presented\cite{Shaklan:88}, taking note again that for the purpose of this study we have ignored real material effects. 


The total throughput within the 5 PL cores exceeds the throughput of the SM fiber for an on-axis PSF over a wide wavelength range upwards of $1.33\upmu$m and remains relatively flat. This displays that when using the PL in configuration (ii) and taking into account all cores, we are essentially using a MM fiber and the total throughput can exceed a SM fiber. This would require accepting multiple SM inputs to an instrument which would, for example, increase the length of a spectrograph slit and thus overall size of a spectrograph. Whether this is a viable option will depend on the specific instrument and whether the disadvantage of increasing the instrument size is worth accepting to gain the PL's ability to remove modal noise.

For a fixed waveguide diameter, shorter wavelengths result in an increased number of guided modes which can also been seen in the Figure~\ref{throughput_plots}. The on-axis SM fiber throughput increases for shorter wavelengths as it is now acting as a MM fiber. The PL's overall throughput decreases however as the shorter wavelengths result in a mode mismatch between the MM and SM sections of the taper and light is therefore lost to the fiber cladding. 

\subsection{Wavefront sensing}
\label{tip/tilt}
\subsubsection{Tip/tilt determination}
\label{tip/tilt determination}

We have assumed an AO-corrected diffraction limited PSF as we expect the PL to be placed in the telescope focal plane after an AO system. Residual image motion at the input of both the SM fiber and PL leads to a reduction in overall throughput of the SM and the central core of the PL. However, when using the PL in configuration (i), analysis of the light coupled into the outer 4 cores of the PL can allow image motion to be measured. 

\begin{figure}[!h]
\hspace*{0.5cm}
\centering\includegraphics[width=13cm]{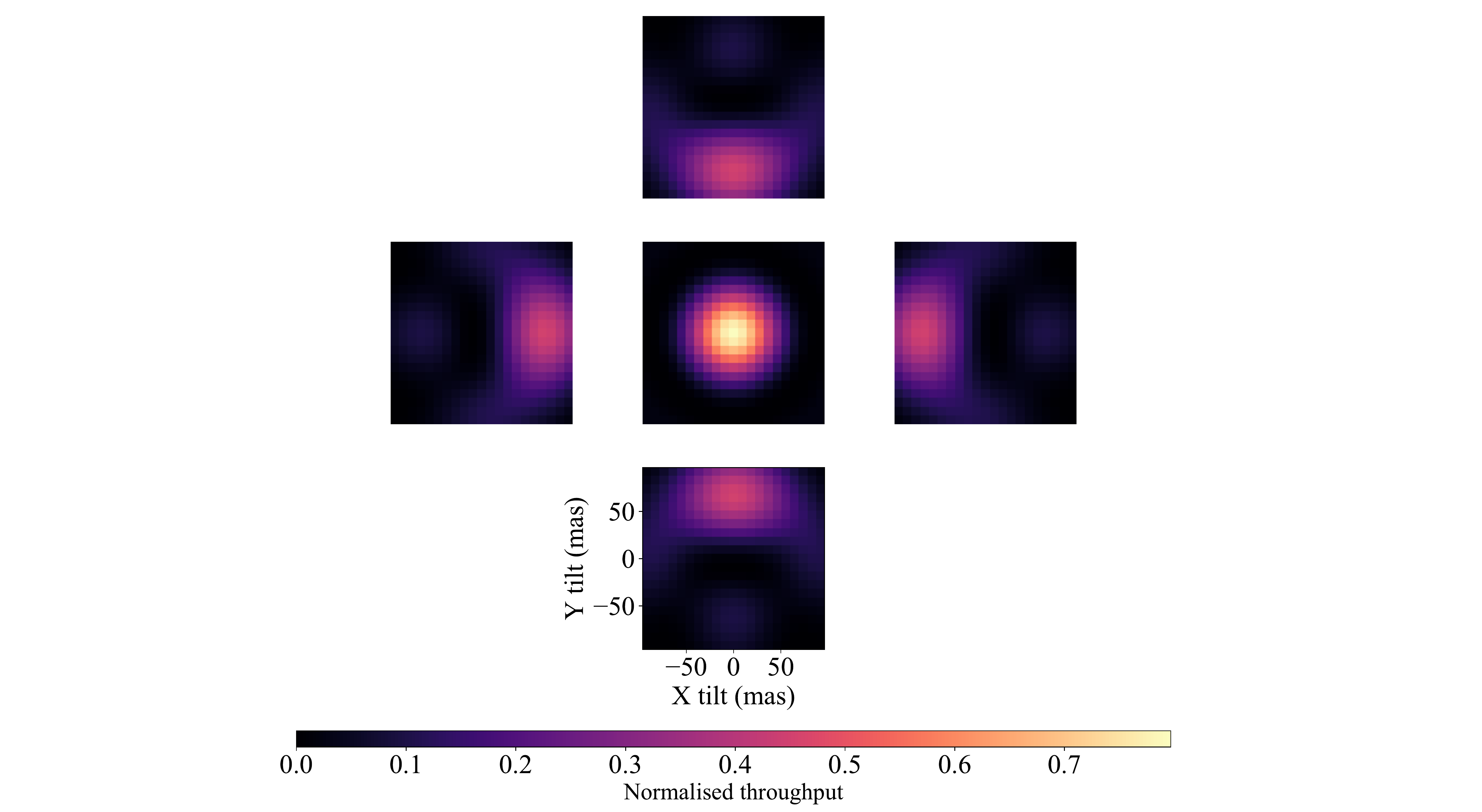}
\caption{Normalized intensity plots for each output core of the PL. The axes show the input x-axis and y-axis tilts in milliarcseconds while the color bar is the normalized intensity value. The position of each plot corresponds to the position of each output at the end face of the PL}\label{all_cores}
\end{figure}

\begin{figure}[!ht]
\centering\includegraphics[width=13cm]{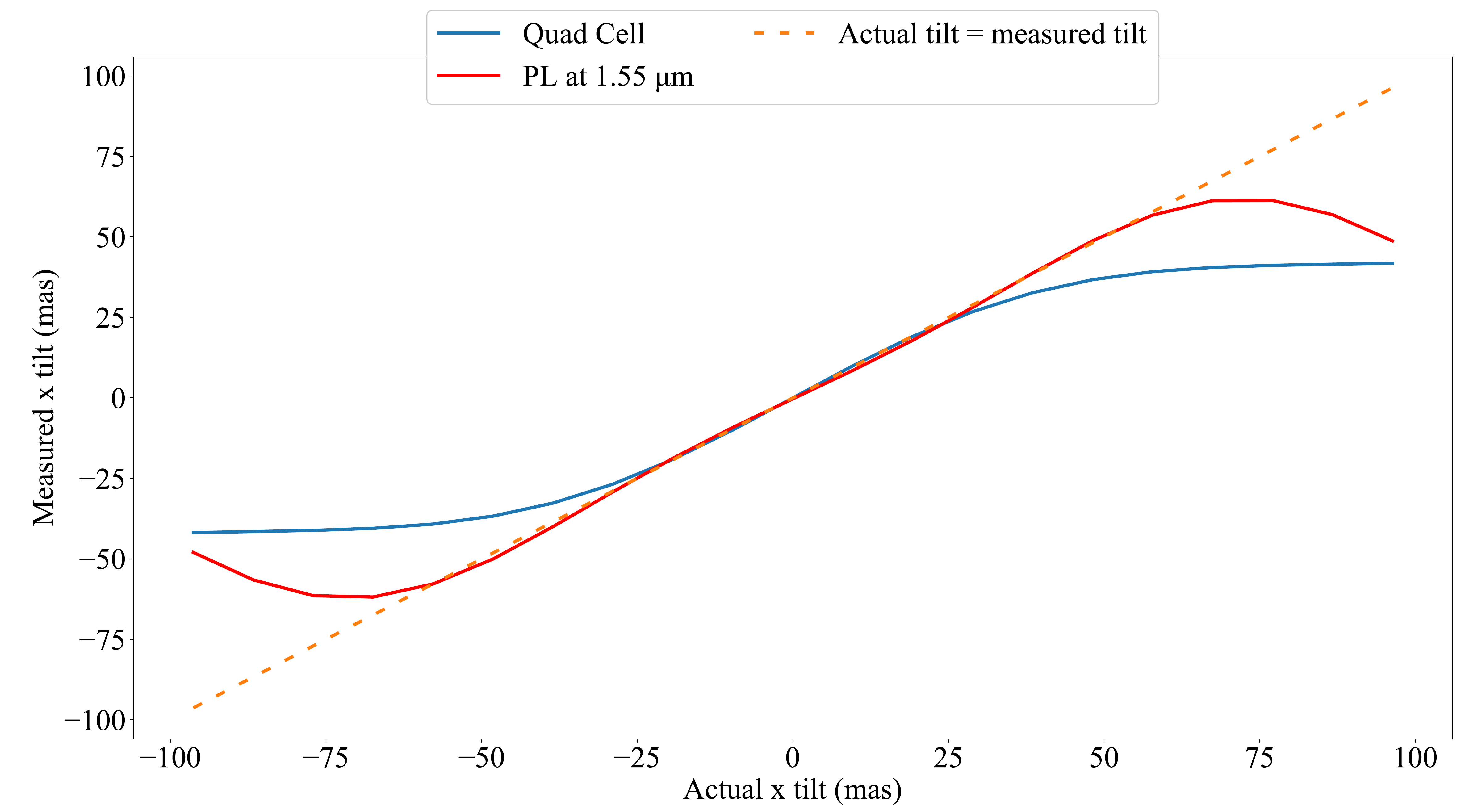}
\caption{Measured x-axis tilt vs. actual x-axis tilt of the input PSF for the PL and quad cell}\label{quad_plot}
\end{figure}

Figure~\ref{all_cores} displays the input PSF position vs. the normalized throughput of each core using $1.55\upmu$m light. It can be seen that the output from each core changes dramatically depending on input PSF position, which allows us to determine the tip/tilt value by measuring the difference in the output core intensities. Figure~\ref{quad_plot} shows the comparison of the PL with a quad cell of detector pixels at measuring the actual input PSF position. The PL displays a linear range of $\pm 55$milliarcseconds (mas) which exceeds that of the quad cell's $\pm 25$mas. The main difference with the PL is that outside of this linear range the measured tilt becomes degenerate, as opposed to the quad cell's signal which saturates at large tilts. For these large tilts which lie outside the PL's linear range, this could cause an issue in a closed loop AO system, as the tilt calculation algorithm could retrieve two different tilt values with the same output intensities from the PL. 

Retrieving tip/tilt measurements when using the PL in configuration (ii) would be achieved in a similar process to configuration (i) described above. The main difference being that each output would need to be separated by wavelength, for example by a spectrograph, to avoid inaccurate results due to the chromatic effects shown in Figure~\ref{throughput_plots}. As a detector in a spectrograph would be running at a much slower rate than that of an AO system, the PL would only be able to make tip/tilt measurements of static or slowly varying NCP aberrations when used in configuration (ii).

\subsubsection{Defocus measurements}

By analyzing the peak wavelength of the central core, or the valley wavelength of the outer cores, we can also determine whether or not the incident PSF on the PL is in focus. The peak wavelength of light through the central core shifts as the input is defocused, which could result in accurate measurements of the defocus term in the input if the output light is analyzed via a spectrograph. Alternatively, if using the device at a single wavelength, the ratio of light in the outer cores with respect to the central core can also be used to make the same measurement. We are currently working on a reconstruction algorithm similar to the tip/tilt determination in section \ref{tip/tilt determination} to investigate if the measurements are linear and importantly, if the results for defocus are orthogonal to the tip/tilt results.

\section{Experimental set up}
\label{experiment}

\begin{figure}[!h]
\hspace*{-0.5cm}
\centering\includegraphics[width=15cm]{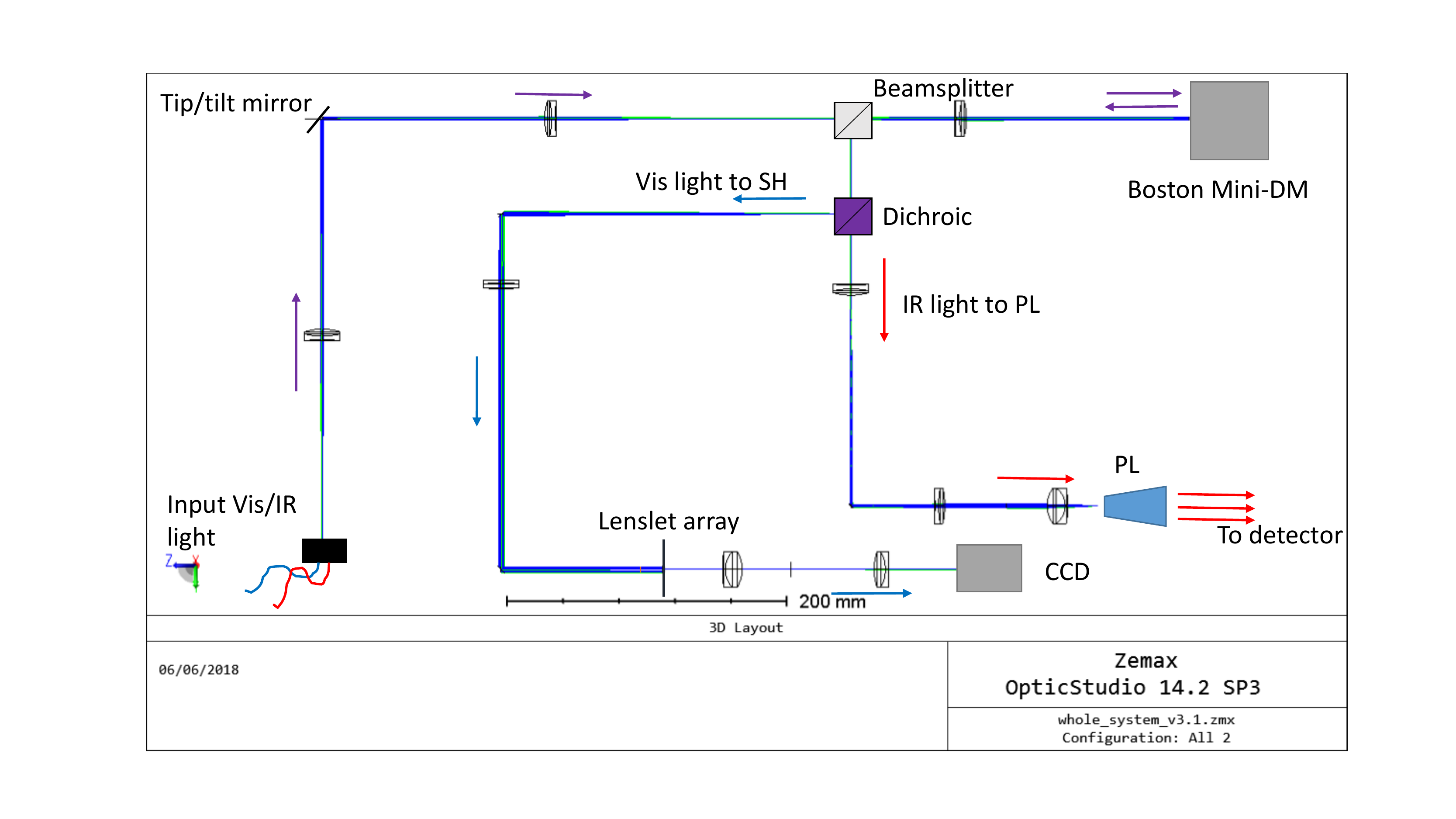}
\caption{Zemax diagram showing the optical set up for the lab experiment.} \label{lab_schematic}
\end{figure}

We have designed an optical test bed to experimentally analyze our optimized PL (Figure~\ref{lab_schematic}). The set up is designed to fit the available space envelope (800mm x 500mm) at the William Herschel 4.2m Telescope (WHT) to allow the possibility of future on-sky testing behind the CANARY instrument. We are using a dual light source of visible ($0.66\upmu$m) and IR ($1.55\upmu$m) for the experiment. The visible light will feed a Shack-Hartmann (SH) WFS while the IR light will be the source for the PL. A tip/tilt mirror will apply the image motion we require to compare the PL with our simulations. 

We also have a Boston Micromachines Mini 32 actuator DM which will be run in closed loop with the SH in the lab to optimize the PSF injected into the PL. The use of a DM also gives the possibility of study into the effects of higher order aberrations in the future.

\section{Conclusions}
We have shown that a 5 core photonic lantern (PL) can be used as a replacement for a single mode (SM) fiber input to an instrument which requires SM resolution. When utilized as a tip/tilt sensor, the PL displays a linear response over a $\pm 55$mas range, which is over double that of a quad cell of detector pixels. The main benefit of this is that as the PL is also the instrument feed, residual and non-common path aberrations can be measured separately to a telescopes adaptive optics (AO) system. 

Configuration (i) can match the throughput of an on-axis SM fiber for the central wavelength of $1.55\upmu$m. The modal beating results in reduced throughput at other wavelengths. However, the ability to stabilize the input could still provide a higher throughput than a SM fiber with no residual aberration correction over wavelength ranges of $1.48$$-$$1.62\upmu$m, for a 0.5 core radii offset of the input PSF, and between $1.40$$-$$1.69\upmu$m, for a 1 core radii input offset.

Configuration (ii) allows the use science data from all cores to retrieve tip/tilt information of slow varying aberrations at the telescope focal plane when used in a spectrograph. This is essentially a MM fiber collecting more light than a SM fiber, but keeping the stability of SM fiber outputs. The drawback of using such a system over a single SM fiber is the resulting increase in size of spectrograph due to the increased number of inputs and the slow tip/tilt measurements, however the elimination of modal noise while collecting as much light as a MM fiber may outweigh this disadvantage.

\section{Acknowledgements}

We would like to thank Sergio Leon-Saval for his valuable insight and time in enabling this simulation work to be undertaken. Mark Corrigan has been funded by STFC under grant ST/M503472/1. Robert J. Harris is supported by the Carl-Zeiss foundation. This project has received funding from the European Union's Horizon 2020 research and innovation programme under grant agreement No 730890. This material reflects only the authors views and the Commission is not liable for any use that may be made of the information contained therein.

\bibliography{report} 
\bibliographystyle{spiebib} 

\end{document}